\documentclass[twocolumn,letterpaper]{IEEEtran}
\usepackage{latexsym}\usepackage{graphicx}
\newcommand{\psfx}[2]{\includegraphics[width=#2]{#1}}
\newcommand{\bfx}{\mbox{\boldmath $x$}}
  
\newcommand{\bfy}{\mbox{\boldmath $y$}}
\newcommand{\bfz}{\mbox{\boldmath $z$}}
\newcommand{\bfr}{\mbox{\boldmath $r$}}
\newcommand{\bfvartheta}{\mbox{\boldmath $\vartheta$}}\newcommand{\bfnu}{\mbox{\boldmath $\nu$}}

\newcommand{\beq}{\begin{equation}}\newcommand{\eeq}{\end{equation}}

\newtheorem{exam}{{\bf Example}}
\usepackage{xfloat}\usepackage{cite} \usepackage{amssymb}\usepackage{amsmath} \DeclareMathOperator*{\argmin}{arg\,min}
\begin{document}
\title{An Unsupervised Learning Approach for Data Detection in the Presence of Channel Mismatch and Additive Noise  
\thanks{Kees A. Schouhamer Immink is with Turing Machines Inc, Willemskade 15d, 3016 DK Rotterdam, The Netherlands. E-mail: immink@turing-machines.com. }
\thanks{Kui Cai is with Singapore University of Technology and Design (SUTD), 8 Somapah Rd, 487372, Singapore. E-mail: cai\_kui@sutd.edu.sg.}
\thanks{This work is supported by Singapore Agency of Science and Technology (A*Star) PSF research grant and SUTD-ZJU grant ZJURP1500102}
}
\author{Kees A. Schouhamer Immink and Kui Cai}
\maketitle
\begin{abstract}  We investigate machine learning based on clustering techniques that are suitable for the detection of encoded strings of $q$-ary symbols transmitted over a noisy channel with partially unknown characteristics. We consider the detection of the $q$-ary data as a classification problem, where objects are recognized from a corrupted vector, which is obtained by an unknown corruption process. We first evaluate the error performance of $k$-means clustering technique without constrained coding. Secondly, we apply constrained codes that create an environment that improves the detection reliability and it allows a wider range of channel uncertainties. \end{abstract}

\section{Introduction} We present new techniques for the detection of $q$-ary data in the face of additive noise and unknown channel corruption by a slow change (drift) of some of the channel parameters. The new detection methods are based on the teachings of cluster analysis. An $n$-symbol $q$-ary word $(x_1,\ldots, x_n)$, $x_i \in \{0,\ldots,q-1\}$ is transmitted or stored, and the received word $(r_1, \ldots, r_n)$ is corrupted by additive noise, intersymbol interference, and other unknown nuisance. Retrieving a replica of the original $q$-ary data is seen as the classification function $(r_1, \ldots, r_n) \rightarrow \{0,\ldots,q-1\}$. Machine learning and deep learning are techniques that are very suitable for classification tasks. The detection function is considered here as a classification problem, or object recognition, which is targeted by cluster analysis. Cluster analysis is an example of unsupervised machine learning, a common technique for statistical data analysis, used in many fields, pattern recognition, image analysis, information retrieval, data compression, and computer graphics~\cite{Goo}. 

We investigate a typical competitive learning algorithm, named {\em $k$-means clustering} technique, which is an iterative process that implements the detection function given initial values of some basic parameters. The aim of the learning algorithm is to map $n$ received symbols into $k$ clusters, where in the case at hand the $k$ clusters are associated with the $q$ symbol values. The detector is ignorant of the number of different symbol values in the sent codeword, that is $k \leq q$. 

A major challenge in cluster analysis is the estimation of the optimal number of `clusters'~\cite{Tib},~\cite{Gor}. The $k$-means clustering technique does not allow to easily estimate the number of (different) clusters, and therefore other means are needed to estimate the number of clusters, $k$. Due to the presence of vexatious codewords and channel distortion, the iteration process may not always converge to a proper solution. To solve this issue, we define constrained coding that may assist in creating an environment where $k$-means clustering technique is a reliable detection technique, and the estimation of the number of clusters can be avoided.   

In mass data storage devices, the user data are translated into physical features that can be either electronic, magnetic, optical, or of other nature~\cite{I52}. Due to process variations, the magnitude of the physical effect may deviate from the nominal values, which may affect the reliable read-out of the data. We may distinguish between two stochastic effects that determine the process variations. On the one hand, we have the unpredictable stochastic process variations, and on the other hand, we may observe long-term effects, also stochastic, due to various physical effects. For example, in non-volatile memories (NVMs), such as floating gate memories, the data is represented by stored charge. The stored charge can leak away from the floating gate through the gate oxide or through the dielectric. The amount of leakage depends on various physical parameters, for example, the device temperature, the magnitude of the charge, the quality of the gate oxide or dielectric, and the time elapsed between writing and reading the data. 

The probability distribution of the recorded features changes over time, and specifically the mean and the variance of the distribution may change. The long-term effects are hard to predict as they depend on, for example, the (average) temperature of the storage device. An increase of the variance over time may be seen as an increase of the noise level of the storage channel, and it has a bearing on the detection quality. The long-term deviations from the nominal means, called {\em offsets}, can be estimated using an aging model, but, clearly, the offsets depend on unpredictable parameters such as temperature, humidity, etc, so that the prediction is inaccurate. 

Various techniques have been advocated for improving the detector resilience in case of channel mismatch when the means and the variance of the recorded features distribution have changed. Estimation of the unknown offsets may be readily achieved by using reference cells, i.e., redundant cells with known stored data. The method is often considered too expensive in terms of redundancy, and alternative methods with lower redundancy have been sought for.

Alternatively, coding techniques can be applied to alleviate the detection in case of channel mismatch. Specifically {\em balanced codes}~\cite{I62},~\cite{Zho},~\cite{Pe6} and {\em composition check codes}~\cite{Sa3},~\cite{I73} preferably in conjunction with Slepian's optimal detection~\cite{Sle} offer resilience in the face of channel mismatch. These coding methods are often considered too expensive in terms of coding hardware and redundancy, specifically when high-speed applications are considered.  

Detectors based on the Pearson distance instead of the traditional Euclidean distance are immune to channel mismatch~\cite{I67}. For the binary case, $q=2$, the redundancy is low and the complexity of the Pearson detector scales with $n$. However, the required number of operations grows exponentially with the length, $n$, and alphabet size, $q$, so that for larger values of $n$ and $q$ the method becomes an impracticability~\cite{I74}. Alternative detection methods for larger $q$ and $n$ that are less costly in resources are welcome. 

In this paper, we investigate detection schemes of $q$-ary, $q>2$, codewords that are based on the results of modern cluster analysis. We assume distortion of the symbols received by additive noise and we further assume that the channel characteristics are not completely known to both sender and receiver. Detection is based on the observation of $n$ symbols only, and the observation of past or future symbols is not assumed.

We set the scene in Section~\ref{secprel} with preliminaries and a description of the mismatched channel model. Prior art detection schemes are discussed in Section~\ref{priorart}. In Section~\ref{kmeans}, we present a new detection based on $k$-means clustering. Computer simulations are conducted to assess the error performance of the prior art and new  schemes developed. In Sections~\ref{secunknownsmall} and ~\ref{secunknownlarge}, we adopt a simple linear channel model where it is assumed that the gain and offset of the received signal are unknown. Computer simulations are conducted to assess the error performance of the detection schemes. Section~\ref{conclus} concludes this paper.
\section{Preliminaries and channel model}\label{secprel}
We consider a communication codebook, ${\cal S} \subseteq {\cal Q}^n$, of selected $n$-symbol codewords $\bfx = (x_1, x_2, \ldots, x_n)$ over the $q$-ary alphabet ${\cal Q} = \{0, \ldots, q-1\}$, where $n$, the {\it length} of $\bfx$, is a positive integer. The codeword, $\bfx \in {\cal S}$, is translated into physical features, where the logical symbols, $i$, are written at an average (physical) level $i+b_i$, where $b_i$ $\in\mathbb{R}$, $0 \leq i \leq q-1$, denotes the average deviation from the nominal or `ideal' value. The average deviations, $b_i$, may slowly vary (drift) in time due to charge leakage or temperature change. The quantities $b_i$ are average deviations, called offsets, from the nominal levels, and they are relatively small with respect to the assumed unity difference (or amplitude) between neighboring physical signal levels. For unambiguous detection, the average of the physical level associated with the logical symbol $`i$' is assumed to be less than that associated with the logical symbol $`i+1$'. In other words, we have the premise
\beq  b_0 < 1+b_1 < 2 + b_2 < \cdots <q-1 + b_{q-1} \label{eqpremise1}\eeq
or
\beq b_{i-1} - b_i < 1, \,\, 1 \leq i \leq q-1 . \label{eqpremise}\eeq 
Assume a codeword, $\bfx$, is sent. The symbols, $r_i$, of the retrieved vector $\bfr = (r_1, \ldots, r_n)$ are distorted by additive noise and given by
\beq r_i = x_i + b_{x_i}  + \nu_i  .  \label{eq_channel}\eeq
We first design a detector for the above case where the unknown offsets, $b_i$'s, are uncorrelated. Thereafter, we distinguish two special cases, where the $b_i$'s are correlated. For the first, general, case, we assume 
\beq  b_i = (a-1)i +b , \eeq 
where $a$ is an unknown attenuation, or {\em gain}, of the channel, and $b$ is an unknown {\em offset}, $a$ and $b$ $\in\mathbb{R}$. We simply find using (\ref{eq_channel}) 
\beq r_i = ax_i + b + \nu_i  .  \label{eq_channel1}\eeq
In the {\em offset-only} case, $a=1$, all $b_i$'s are equal, or $r_i = x_i + b + \nu_i$. We assume that the received vector, $\bfr$, is corrupted by additive Gaussian noise $\bfnu$ = $(\nu_1, \ldots, \nu_n)$, where $\nu_i \in \mathbb{R}$ are zero-mean independent and identically distributed (i.i.d) noise samples with normal distribution ${\cal N}(0,\sigma^2)$. The quantity $\sigma^2\in \mathbb{R}$ denotes the noise variance. The additive noise term may be caused by fabrication process variations or electronics (detector) noise. 
\section{Prior art detection schemes}\label{priorart}
Below we discuss three prior art detection schemes and relevant properties.
\subsection{Fixed threshold detection (FTD)}\label{secthreshold}
The symbols of the received word, $r_i$, can be straightforwardly quantized to an integer, $\hat x_i\in{\cal Q}$, with a conventional {\em fixed threshold detector} (FTD), also called {\em symbol-by-symbol detector}. The threshold function is denoted by $\hat x_i = \Phi_{\vartheta}(r_i)$, $\hat x_i\in{\cal Q}$, where the threshold vector $\bfvartheta =$ $(\vartheta_0, \ldots, \vartheta_{q-2})$ has  $q-1$ (real) elements, called {\em thresholds} or {\em threshold levels}. The threshold vector satisfies the order 
\beq \vartheta_0 < \vartheta_1 < \vartheta_2 < \cdots < \vartheta_{q-2}.  \eeq
The quantization function, $\Phi_{\vartheta}(u)$, of the threshold detector is defined by
\beq \Phi_{\vartheta}(u) = \left \{  \begin{tabular}{ll}  
0, & $u < \vartheta_0$,\\
$i$, & $\vartheta_{i-1} \leq u < \vartheta_i, 1\leq i \leq q-2,$\\
$q-1$,& $u \geq \vartheta_{q-2}$. \\
\end{tabular} \right . \label{eqQu}\eeq
For a {\em fixed} threshold detector the $q-1$ detection thresholds values, $\vartheta_i$, are equidistant at the levels 
\beq \vartheta_i = \frac{1}{2}+i, \,\,\, 0 \leq i \leq q-2. \eeq
Threshold detection is very attractive for its implementation simplicity. However, the error performance seriously degrades in the face of channel mismatch~\cite{I67}. A detector that dynamically adjusts the thresholds is an alternative that offers solace in the face of channel mismatch. The next subsection describes a typical example. 
\subsection{Dynamic threshold detection (min-max detector)}\label{secdtd}
We assume that the channel model is, see (\ref{eq_channel1}), $r_i=ax_i+b+\nu_i$, where the gain, $a>0$, and offset, $b$, are unknown parameters, except for the sign of $a$. In case ${\cal S}={\cal Q}^n$, that is all possible codewords are allowed, mismatch immune detection is not possible since such a detector cannot distinguish between the word $\hat\bfx$ and its shifted and scaled version $\hat\bfy=c_1\hat\bfx+c_2$. A designer must judiciously select codewords from ${\cal Q}^n$ given adequate constraints that may enable mismatch immune detection. For example, we select for ${\cal S}$ those codewords where the symbols `0' and `$q-1$' must be both at least once present. For the binary case, $q=2$, this implies a slight redundancy as only the all-1 and all-0 words have to be removed,  see Subsection~\ref{secconcod} for details. Then, the detector can straightforwardly estimate the gain and offset by 
\beq \hat a = \frac{\max_i r_i - \min_i r_i}{q-1}  \eeq  
and 
\beq \hat b = \min_i {r_i}, \eeq  
where $\hat a$ and $\hat b$ denote the estimates of the actual channel gain and offset~\cite{I66}. The dynamic thresholds, denoted by $\hat\vartheta_i$, are scaled in a similar fashion as the received codeword, that is,
\beq  \hat \vartheta_i = \hat a \vartheta_i + \hat b, \,\,\, 0 \leq i \leq q-2 . \label{eqvartheta} \eeq 
It has been shown~\cite{I66} that the min-max detector operates over a large range of unknown parameters $a$ and $b$. However, since the estimates, $\hat a$ and $\hat b$, are biased, the above dynamic threshold detector loses error performance with respect to the matched case, especially for larger codeword length $n$. The detector complexity scales linearly with $n$ as the principal cost is the finding of the maximum and minimum of the $n$ received symbol values. Alternatively, detection based on the prior art Pearson distance, discussed in the next subsection, improves the error performance, but with mounting hardware requirements.   
\subsection{Pearson distance detection}\label{secpearson}
Immink and Weber~\cite{I67} advocated the Pearson distance instead of the conventional Euclidean distance for improving the error performance of a mismatched noisy channel. We first define two quantities, namely the {\it vector average} of the $n$-vector ${\bfz}$
\beq \overline{z} =  \frac{1}{n} \sum_{i=1}^n z_i \label{eqms6b}\eeq
and the (unnormalized) {\it vector variance} of $\bfz$
\beq \sigma^2_{z} =  \sum_{i=1}^n (z_i - \overline{z})^2 \label{eqms6a}.  \eeq
The Pearson distance, $\delta_\text{p}(\bfr, \hat\bfx)$, between the received vector $\bfr$ and a codeword $\hat\bfx \in {\cal S}$ is defined by
\beq  \delta_\text{p}(\bfr, \hat{\bfx}) = 1 - \rho_{ \bfr, \hat{\bfx}} , \label{eqms6c} \eeq
where 
\beq \rho_{ \bfr, \hat\bfx} = \frac{ \sum_{i=1}^n (r_i - \overline {r})(\hat x_i - \overline {\hat x} )} {\sigma_r \sigma_{\hat x } } \label{eqms6cc} \eeq
is the well-known {\it (Pearson) correlation coefficient}. It is assumed that both codewords $\bfx$ and $\hat{\bfx}$ are taken from a judiciously chosen codebook $S$, whose properties are explained in subsection~\ref{secconcod}. The Pearson distance is not a metric in the strict mathematical sense, but in engineering parlance it is still called a 'distance' since it provides a useful measure of similarity between vectors. A minimum Pearson distance detector outputs the codeword
\beq {\bfx_\text{o}} = \argmin_{\hat{\bfx}\in S} \delta_\text{p} (\bfr, \hat\bfx) .  \label{eqms1} \eeq
It can easily be verified that the minimization of $\delta_\text{p}(\bfr, \hat{\bfx})$, and thus ${\bfx_\text{o}}$, is independent of both $a$ and $b$, so that the detection quality is immune to unknown drift of the quantities $a$ and $b$. The minimization operation (\ref{eqms1}) requires $|S|$ computations, which is impractical for larger $S$. The number of computations can be reduced to $K$, the number of constant composition codes that constitutes the codebook $S$, given by~\cite{I72}
\beq K = \binom {n+q-3}{q-1} .\label{eqncomp} \eeq 
For the binary case, $q=2$, we have $K=n-1$, so that the detection algorithm (\ref{eqms1}) scales linearly with $n$. For the non-binary case, it is hard to compute or simulate the error performance of minimum Pearson distance detection in a relevant range of $q$ and $n$ as the number, $K$, of operations grows rapidly with both $q$ and $n$. For example, for $q=4$ and $n=64$ we have $K=43.680$ comparisons (\ref{eqms1}) per decoded codeword.

The three prior art detection methods discussed above have drawbacks in error performance and/or complexity, and to alleviate these drawbacks, viable alternatives are sought for. In the next section, we propose and investigate a novel detection method with less complexity requirements, which is based on clustering techniques.
\section{Data detection using $k$-means clustering}\label{kmeans}
In the next subsection we describe the basic $k$-means clustering algorithm, and present results of simulations for the unmatched noisy channel. 
\subsection{Basic $k$-means clustering algorithm}
The $k$-means clustering technique aims to partition the $n$ received symbols into $k$ sets $V = \{V_0, V_1, \ldots , V_{k-1}\}$ so as to minimize the within-cluster sum of squares defined by
\beq \argmin_V \sum_{i=0}^{k-1} \sum_{r_j \in V_i} (r_j-\mu_i)^2 , \label{eqcluster0} \eeq
where the {\em centroid} $\mu_i$ is the mean of the received symbols in cluster $V_i$, or
\beq \mu_i  = \frac{1}{|V_i|} \sum_{r_j \in V_i} r_j . \eeq 
The problem of choosing the correct number of clusters is hard, and numerous prior art publications are available to facilitate this choice~\cite{Tib},~\cite{Gor}. Here we assume that a cluster is associated with one of the $k$ symbol values, that is $k=q$. 

The $k$-means clustering algorithm is an iteration process that finds a solution of (\ref{eqcluster0}). The initial sets $V^{(1)}_i$, $0 \leq i \leq k-1$, are empty. The superscript integer in parentheses, $(t)$, denotes the iteration tally. We initialize the $k$ centroids $\mu^{(1)}_i$, $0 \leq i \leq k-1$, by a reasonable choice. For example, Forgy's method~\cite{For} randomly chooses $k$ symbols (assuming $k<n$), $r_i$, from the received vector $\bfr$, and uses these as the initial centroids $\mu^{(1)}_i$, $ 1 \leq i \leq k$. The choice of the initial centroids has a significant bearing on the error performance of the clustering detection technique. We do not follow Forgy's approach, and try, dependent on the specific situation at hand, to develop more suitable initial centroids $\mu^{(1)}_i$'s. We assume that we order the centroids such that 
\beq 
\mu_0^{(t)} < \mu_1^{(t)} < \cdots < \mu_{q-1}^{(t)} . \label{eqcent}
\eeq
After the initialization step, we iterate the next two steps until the symbol assignments no longer change. 
\begin{itemize}
\item {{\bf Assignment step:} Assign the $n$ received symbols, $r_i$, to the $k$ sets $V^{(t+1)}_j$. If $r_i$, $1 \leq i \leq n$, is closest to $\mu^{(t)}_{\ell}$, or
\beq \ell = \argmin_j \left (r_i -\mu^{(t)}_j \right)^2 , \label{eqassign}\eeq
then $r_i$ is assigned to $V^{(t+1)}_{\ell}$.} The (temporary) decoded codeword, denoted by 
\beq \hat\bfx^{(t)} = (\hat x^{(t)}_1, \ldots, \hat x_n^{(t)}), \label{eqhatx} \eeq
is found by 
\beq  \hat x_{i}^{(t)} = \phi_{V^{(t)}} (r_i) , \,\, 1 \leq i \leq n, \eeq
where $\phi_{V^{(t)}}(r_i) = j$ such that $r_i \in V_j^{(t)}$. 
\item {{\bf Updating step:} Compute updated versions of the $k$ means $\mu^{(t+1)}_j, j \in {\cal Q}$. An update of the new means $\mu^{(t+1)}_j$ is found by
\beq \mu^{(t+1)}_j = \frac{1}{|V^{(t+1)}_j|} \sum_{r_i \in V^{(t+1)}_j} r_i, \,\, j \in {\cal Q} , \label{equpdate}\eeq 
where it is understood that if $|V^{(t+1)}_j|=0$ that $\mu^{(t+1)}_j=\mu^{(t)}_j$ (that is, no update). 
}\end{itemize}
After running the above routine until the temporary decoded word is unchanged, say at iteration step, $t=t_o$, 
we have $\hat\bfx^{(t_o-1)}=\hat\bfx^{(t_o)} $. Then we have found the final estimate of the sent codeword, 
$\bfx_o=\hat\bfx^{(t_o)}$. Bottou~\cite{Bot} showed that the $k$-means cluster algorithm always converges to a simple steady state, 
and limit cycles do not occur. It is possible, however, that the process reaches a local minimum of the within-cluster sum of squares (\ref{eqcluster0}).
\subsection{Assignment step: relation with threshold detection}
We take a closer look at the assignment step of the $k$-means clustering technique, given by (\ref{eqassign}). Considering the order (\ref{eqcent}) of the centroids $\mu^{(t)}_j$, we simply infer that the symbol $r_i$ lies between, say, $\mu^{(t)}_u \leq r_i \leq \mu^{(t)}_{u+1}$, $ 0 \leq u \leq q-2$. Thus
\beq \ell = \argmin_{j\in \{u,u+1\}} \left (r_i -\mu^{(t)}_j \right)^2 . \nonumber  \eeq
As
\begin{eqnarray}  &&\left (r_i -\mu^{(t)}_u\right )^2 -  \left (r_i -\mu^{(t)}_{u+1} \right )^2 \nonumber\\ 
&=& 2r_i \left(\mu^{(t)}_{u+1}-\mu^{(t)}_u \right) + \left (\mu^{(t)}_u \right)^2- \left(\mu^{(t)}_{u+1} \right)^2 \nonumber\\ 
&=& \left(\mu^{(t)}_{u+1}-\mu^{(t)}_u \right) \left( 2r_i-\mu^{(t)}_u -\mu^{(t)}_{u+1}  \right) ,
\end{eqnarray}
we obtain
\beq \ell = \left \{ 
\begin{tabular}{ll}  $u$, & $r_i < \frac{\mu^{(t)}_{u+1}+\mu^{(t)}_u} {2}$,\\ $u+1$, & $r_i > \frac{\mu^{(t)}_{u+1}+\mu^{(t)}_u} {2}$ .\\
\end{tabular} \right .\eeq
Using (\ref{eqQu}), we yield
\beq \ell = \argmin_j \left (r_i -\mu^{(t)}_j \right)^2 = \Phi_{\hat \vartheta}(r_i),\eeq
where the threshold vector, $\hat\bfvartheta$, is given by
\beq \hat \vartheta_i = \frac {\mu^{(t)}_{i+1} + \mu^{(t)}_i}{2}, \,\, 0 \leq i \leq q-2,  \label{eqvartheta1} \eeq 
and the intermediate decoded vector, $\hat\bfx^{(t)}$, is given by 
\beq  \hat x_i^{(t)} = \Phi_{\hat\vartheta}(r_i) , \,\, 1 \leq i \leq n . \eeq 
We conclude that the $k$-means cluster detection method is a dynamic threshold detector, where at each update the threshold vector, $\hat\bfvartheta$, is updated with the means of the members of each cluster using~(\ref{equpdate}).

In the next section, we report on outcomes of computer simulations using channel model~(\ref{eq_channel}).
\subsection{Results of simulations}\label{simres}
We investigate the error performance of channel model (\ref{eq_channel}), where we assume that the stochastic deviations from the means, $b_i$, $i \in {\cal Q}$, are taken from a zero-mean continuous uniform distribution with variance $\sigma^2_b$. Thus, the $b_i$'s lie within the range $-\sqrt{3}\sigma_b \leq b_i \leq \sqrt{3}\sigma_b$. We assume a uniform distribution to guarantee premise (\ref{eqpremise}). 

\begin{figure} \centerline{\psfx{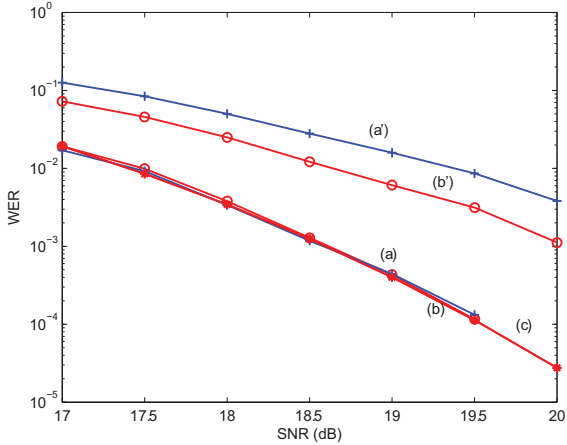}{8.5cm}} \caption{\protect\small Word error rate (WER) of fixed threshold detection (FTD), curve (a'), and $k$-means clustering detection, curve (b'), versus SNR $=-20 \log \sigma$ (dB) for $n=64$, $q=4$, and $\sigma_b=0.1$. Curves (a) and (b) are shown for the case $\sigma_b=0$ (ideal channel). The upperbound (\ref{equpbound}) to the word error rate of a fixed threshold detector for an ideal noisy channel, $q=4$ and $n=64$, curve (c). \label{figclus}} \end{figure}

We simply initialize the centroids by $\mu^{(1)}_i=i$, $i \in {\cal Q}$, and iterate the assignment and updating steps as outlined above. Figure~\ref{figclus} shows outcomes of computer simulations for the case $n=64$ and $q=4$, where we compare the word error rate (WER) of conventional fixed threshold detection and the novel dynamic threshold detection based on $k$-means clustering classification versus the signal-to-noise ratio (SNR) defined by $=-20 \log \sigma$. We plotted two cases, namely $\sigma_b=0$ (ideal channel) and $\sigma_b=0.1$. As a further comparison we plotted the upper bound of the word error rate of a threshold detector for an ideal additive noise channel, given by~\cite{I67} 
\beq  {\rm WER } < \frac{2(q-1)}{q} n Q \left( \frac {1}{2\sigma} \right) . \label{equpbound} \eeq 
We infer that in case the channel is ideal, $\sigma_b=0$, that the error performance of $k$-means clustering detection is close to the performance of both theory and simulation practice of conventional fixed threshold detection. In case the channel is not ideal, $\sigma_b=0.1$, $k$-means clustering detection is superior to fixed threshold detection. 

The number of iterations, which is an important (time) complexity issue, depends on the integers $q$, $n$, and the signal-to-noise ratio, SNR. The convergence of the iteration process is guaranteed~\cite{Bot}, but the speed of convergence is an open question that we studied by computer simulations.
\begin{table}\caption{Histogram of the number of iterations for $q=4$, $n=64$, and $\sigma_b=0.1$.}$$\begin{array}{l|rr} \hline 
t_o  & \mbox{SNR}=17 \mbox{ dB}& \mbox{SNR}=20\mbox{ dB}\\\hline 1 & 91.43 & 99.70\\2 & 8.37 & 0.30\\3 & 0.20 & 0\\ \hline \end{array}$$ \label{tabiter} \end{table} 
Table~\ref{tabiter} shows results of simulations for the case $q=4$, $n=64$, and $\sigma_b=0.1$ (same parameters as used in the simulations depicted in Figure~\ref{figclus}). At an SNR = 17~dB, around $91 \%$ of the received words is detected without further iterations. In $8 \%$ of the detected words, only one iteration of the threshold levels is needed. At an SNR = 20~dB, we found that all but no iterations are required. Thus, since in the large majority of cases no iterations are needed, we conclude that at the cost of a slight additional (time) complexity, the proposed $k$-means clustering classification outperforms fixed threshold detection. 

\section{Unknown gain $a$ and offset $b$ (small range of uncertainty)}\label{secunknownsmall}
\begin{figure} \centerline{\psfx{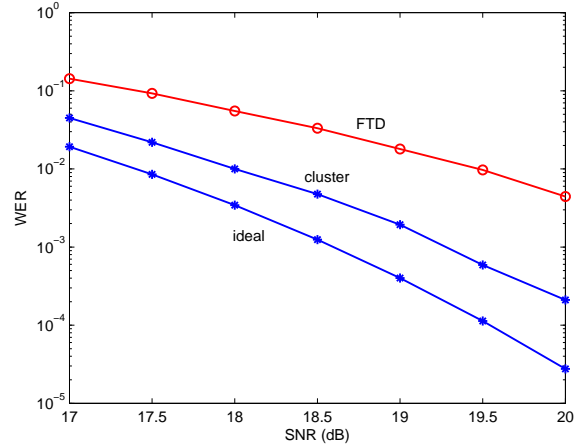}{8.5cm}} \caption{\protect\small Word error rate (WER) of fixed threshold detection (FTD) and $k$-means clustering detection (cluster) versus SNR for $n=64$, $q=4$, $a=0.95$, and $b=0$. As a reference, the upperbound to the word error rate of a fixed threshold detector for the ideal noisy channel given by (\ref{equpbound}).\label{figclus2}}\end{figure}

In this section, we assume that the linear channel model, see (\ref{eq_channel1}), $r_i = ax_i + b + \nu_i$, applies. In case the gain, $a$, is within a tolerance range close to unity and the tolerance range of the offset, $b$, is close to zero, we may directly apply the basic $k$-means clustering as outlined in the previous section. We require that both $a$ and $b$ are so close to their nominal values that a fixed threshold detector works correctly in the noiseless case. Then, the initialization, using the fixed threshold detector, furnishes sufficiently reliable data for the iterations to follow. From the definition of a fixed threshold decoder, see (\ref{eqQu}), we simply derive the following tolerance ranges of $a$ and $b$ that guarantee a flawlessly operating threshold detector, namely 
\begin{eqnarray}b &<& \vartheta_0 =\frac{1}{2} ,\nonumber\\
\vartheta_{i-1} &<& ai +b < \vartheta_i,\,\, 1 \leq i \leq q-2,\nonumber\\
a(q-1)+b &>& \vartheta_{q-2} = q-\frac{3}{2}, \end{eqnarray}
or
\begin{eqnarray}b &<& \frac{1}{2} ,\nonumber\\
i-\frac{1}{2} &<& ai +b < i+\frac{1}{2},\,\, 1 \leq i \leq q-2,\nonumber\\
a(q-1)+b &>& q- \frac{3}{2} \label{eqtolerance}. \end{eqnarray}
Figure~\ref{figclus2} shows outcomes of computer simulations, where we compare for the case $n=64$ and $q=4$, the word error rate (WER) of fixed threshold detection and detection based on $k$-means clustering versus the signal-to-noise ratio (SNR), where the channel gain equals $a=0.95$ and $b=0$. 
We conclude that the cluster detector shows a greater resilience in the face of unknown gain, $a$, and additive noise than the fixed threshold detector. 

In the above case, the parameters  $a$ and $b$ are assumed to have a limited range of uncertainty. In case, however, they have a wider tolerance range than prescribed by (\ref{eqtolerance}), it is not possible to unambiguously detect the codeword with a fixed threshold detector. The detector needs assistance, and constrained coding is applied to assist in overcoming this difficulty as discussed in the next section.
\section{Unknown gain $a$ and offset $b$ (large range of uncertainty)}\label{secunknownlarge}
In this section, we focus on the situation where we anticipate that both parameters $a$ and $b$ have such a great range of possible values that a fixed threshold detector fails in the majority of cases, even in the noiseless case. In the next subsection, we show, by example, that in such a case it is impossible to distinguish between certain nettlesome situations, and constrained coding becomes a requirement to solve the ambiguity. 
\subsection{Constrained coding}\label{secconcod}
In order to cope with larger uncertainties of both parameters $a$ and $b$, we face an ambiguity problem. For example, let $q=5$, and let $(2,4,4)$ be the received vector. Clearly, it is impossible to distinguish between the two choices, where the sent codeword is $(2,4,4)$ and $a=1$ or where $(1,2,2)$ and $a=2$. Let ${\cal S}$ be the adopted codebook, then we can cope with the above ambiguity if $(2,4,4) \in {\cal S}$ then $(1,2,2) \notin {\cal S}$, or {\em vice versa}. The name {\it Pearson code} was coined for a set of codewords that can be uniquely decoded by a detector immune to large uncertainties in both $a>0$ and $b$~\cite{I66}. Codewords in a Pearson code, ${\cal S}$, satisfy two conditions, namely
\begin{itemize}
\item {\it Property A:} If $\bfx \in {\cal S}$ then $c_1+c_2 \bfx \notin {\cal S}$ for all $c_1,c_2\in \mathbb{R}$ with $(c_1,c_2)\ne (0,1)$ and $c_2>0$.
\item {\it Property B:} $\bfx = (c,c,\ldots,c) \notin {\cal S}$ for all $c\in \mathbb{R}$.
\end{itemize}
We adopt a Pearson code that has codewords with at least one `0' symbol and at least one `$q-1$' symbol. We may easily verify that such codewords satisfy Properties~A and B. The number of allowable $n$-symbol codewords equals~\cite{I66}
\beq |{\cal S}| = q^n- 2(q-1)^n +(q-2)^n,  q>1. \label{eqn2}   \eeq
For the binary case, $q=2$, we simply find that
$$|{\cal S}|= 2^n-2$$ 
(both the  all-`1' and all-`0' words are deleted). 
\subsection{Revised $k$-means clustering using min-max initialization}
Here it is assumed that the parameters $a$ and $b$ are completely unknown, except for the sign of $a$, $a>0$. Due to the large uncertainty, we cannot adopt the elementary choice of the initial values of the centroids $\mu^{(1)}_i$ as described in Section~\ref{kmeans}. We propose, following the min-max detector technique described in Subsection~\ref{secdtd}, the choice of the initial centroids $\mu^{(1)}_i$'s using the minimum, $\min_i r_i$, and maximum value, $\max_i r_i$, of the received symbols. The Pearson code guarantees at least one `0' symbol and also at least one `$q-1$' symbol in a codeword. The detector may therefore use the minimum and maximum value of the received symbols as anchor points defining the range of values of the symbols in the received vector. To that end, let 
\beq \alpha_0 = \min_i r_i \mbox{ and } \alpha_1 = \max_i r_i. \label{eqmaxmin} \eeq
The $q$ initial centroids, $\mu^{(1)}_i$, are found by the interpolation
\beq \mu^{(1)}_i = \alpha_0 + (\alpha_1-\alpha_0) \frac{i}{q-1}  , \,\, 0 \leq i \leq q-1 .\eeq  
\begin{figure} \centerline{\psfx{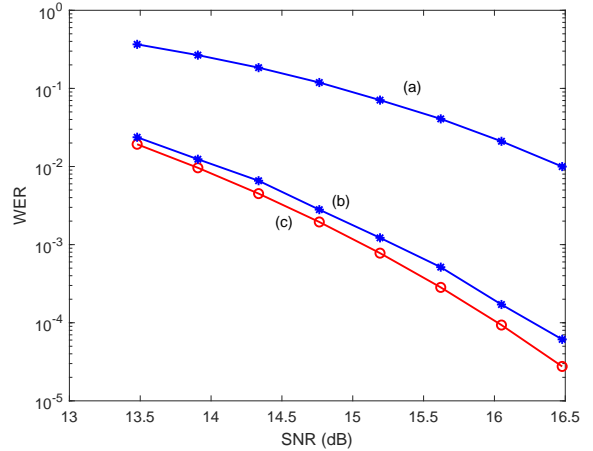}{8.5cm}} \caption{\protect\small Word error rate (WER) for the case $q=4$, $n=64$ and gain $a=1.5$ of a) prior art min-max detector as described in Subsection~\ref{secdtd}, b) $k$-means clustering algorithm, and c) upperbound (\ref{equpbound}) of an ideal fixed threshold detector. Note that the signal-to-noise ratio is defined by ${\rm SNR} =-20 \log (\sigma/a)$. The error performance is independent of the offset $b$.\label{figwer9}} \end{figure}
Note that the above initialization step of the modified $k$-means clustering technique has the same effect as the scaling used in the min-max detector (\ref{eqvartheta}). Figure~\ref{figwer9} shows results of computer simulations for the case $q=4$ and $n=64$ and a gain $a=1.5$. For normalization purposes, we define the SNR by ${\rm SNR} =-20 \log (\sigma/a)$. We compared prior art DTD with the $k$-means clustering detection algorithm. The detector based on $k$-means clustering outperforms the prior art min-max detector. 

In the next subsection, we discuss a second modification to the basic $k$-means clustering method using regression analysis.  
\subsection{Revised $k$-means clustering algorithm using regression analysis}\label{secrevkmeans}
We adopt a second modification to the clustering algorithm of Section~\ref{kmeans}. In the basic updating step (\ref{equpdate}), the $k$ clusters centroids are updated by computing a new mean of the members in that cluster only. Here we assume that the linear channel model, $r_i = ax_i + b + \nu_i$, described by (\ref{eq_channel1}) holds. We have investigated an alternative method for updating the centroids, $\mu^{(t+1)}_j$, by applying the well-known {\em linear regression model}~\cite{Hog} that estimates the two coefficients $a$ and $b$ instead of the $q$ centroids $\mu_i$. 

\begin{figure} \centerline{\psfx{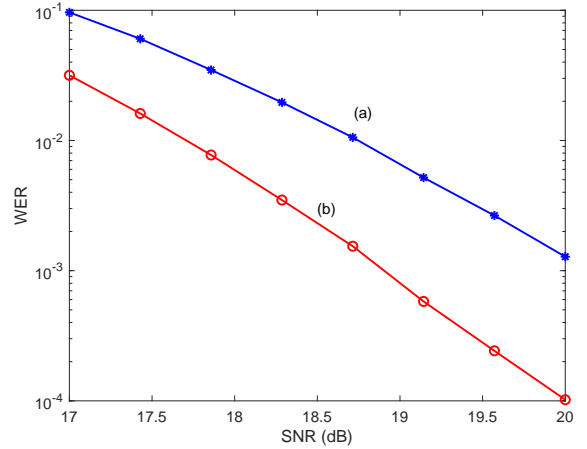}{8.5cm}} \caption{\protect\small Word error rate (WER) of a) revised k-means clustering using min-max initialization, and b) revised $k$-means clustering algorithm using regression method for the case for $q=16$ and $n=64$. \label{figwer15}} \end{figure}

We start and initialize as described in the previous subsection, where the $q$ initial centroids, $\mu^{(1)}_i$, are found by the interpolation
\beq \mu^{(1)}_i = \alpha_0 + (\alpha_1-\alpha_0)i  , \,\, 0 \leq i \leq q-1 ,\eeq  
where, as in (\ref{eqmaxmin}),
\beq \alpha_0 = \min_i r_i \mbox{ and } \alpha_1 = \max_i r_i.  \eeq
For the offset only case, $a$=1, we have 
\beq \mu^{(1)}_i = \alpha_0 + i  , \,\, \,\, i=0, \ldots, q-1. \eeq  
After the initialization, we iterate the next two steps until equilibrium is reached. 
\begin{itemize}
\item {{\bf Assignment step:} Assign the $n$ received symbols, $r_i$, to the $k$ sets $V^{(t+1)}_j$. If $r_i$, $1 \leq i \leq n$, is closest to $\mu^{(t)}_{\ell}$, or
\beq \ell = \argmin_j \left (r_i -\mu^{(t)}_j \right)^2 , \eeq
then $r_i$ is assigned to $V^{(t+1)}_{\ell}$.} The (temporary) decoded codeword, denoted by 
\beq \hat\bfx^{(t)} = (\hat x^{(t)}_1, \ldots, \hat x_n^{(t)}), \eeq
is found by 
\beq \hat x_{i}^{(t)} = \phi_{V^{(t)}} (r_i) , \,\, 1 \leq i \leq n, \eeq
where $\phi_{V^{(t)}}(r_i)=j$ such that $r_i \in V_j^{(t)}$. 
\item {{\bf Updating step:} Updates of the means $\mu^{(t+1)}_j, j \in {\cal Q}$ are found by a linear regression model that estimates the coefficients $a$ and $b$. To that end, define the linear regression model 
\beq \hat r_i = \hat a^{(t)} \hat x_i^{(t)} + \hat b^{(t)} , \eeq 
where the (real-valued) {\em regression coefficients} $\hat a^{(t)}$ and $\hat b^{(t)}$, chosen to minimize $\sum_{i=1}^n(r_i-\hat r_i)^2$, denote the estimates of the unknown quantities $a$ and $b$. The regression coefficients $\hat a^{(t)}$ and $\hat b^{(t)}$ are found by invoking the well-known linear regression method~\cite{Hog}, and we find using (\ref{eqms6a}) and (\ref{eqms6cc}) 

\beq \hat a^{(t)} = \frac{ \sum_{i=1}^n (r_i - \overline {r})(\hat x_i - \overline {\hat x} )} {\sum_{i=1}^n (\hat x_i - \overline{\hat x})^2 }  = \frac{\sigma_r}{\sigma_{\hat x^{(t)}} } \rho_{\bfr, \hat \bfx^{(t)}}   \label{eqestb} \eeq
and
\beq \hat b^{(t)} = \overline r - \hat a^{(t)} \overline{\hat x^{(t)}} . \label{eqesta} \eeq 
We note that for all $\bfx \in S$, $\sigma_{\hat x^{(t)}} \neq 0$ since Property~B holds, see Subsection~\ref{secconcod}. The updated $\mu^{(t+1)}_i$, $i=0, \ldots, q-1$, are found by the interpolation  
\beq \mu^{(t+1)}_i  = \hat a^{(t)} i + \hat b^{(t)} .  \label{eqmut1}\eeq 
For the offset-only case, $a=1$, we simply find
\beq  \hat b^{(t)} = \overline r - \overline{\hat x^{(t)}} ,\eeq
and
\beq \mu^{(t+1)}_i  = i + \hat b^{(t)} = i + \overline r - \overline{\hat x^{(t)}} .  \eeq 
}\end{itemize}
We have conducted a myriad of computer simulations with the above algorithms. Figure~\ref{figwer15} compares the error performance of the revised $k$-means clustering using min-max initialization versus the revised $k$-means clustering using regression analysis for the case $q=16$ and $n=64$. The performance difference between the two cluster methods is independent of the unknown quantities $a$ and $b$. 
\section{Conclusions}\label{conclus}
We have proposed and analyzed machine learning based on a $k$-means clustering technique as a detection method of encoded strings of $q$-ary symbols. We have analyzed the detection of distorted data retrieved from a data storage medium where user data is stored as physical features with $q$ different levels. Due to manufacturing tolerances and ageing the $q$ levels differ from the desired, nominal, ones. Results of simulations have been presented, where the $q$ unknown level differences, called offsets, are independent stochastic variables with a uniform probability distribution. We have evaluated the error performance of $k$-means clustering detection technique, where the offsets are correlated, and can be modelled as unknown scale, or gain, and translation, or offset. At the cost of some additional (time) complexity, the proposed $k$-means clustering classification outperforms common prior art dynamic detection methods in the face of additive noise and channel mismatch. 

\end{document}